\documentclass[conference]{IEEEtran}
\IEEEoverridecommandlockouts

\usepackage{cite}
\usepackage{amsmath,amssymb,amsfonts}
\usepackage{algorithmic}
\usepackage{microtype}%
\usepackage{graphicx}
\usepackage{textcomp}
\usepackage{xcolor}
\usepackage{mathtools}
\usepackage{multirow}
\usepackage{placeins}
\usepackage{subcaption}
\usepackage{tikz}
\usepackage{url}
\usepackage{listings}%
\usepackage[linesnumbered]{algorithm2e}%

\def\BibTeX{{\rm B\kern-.05em{\sc i\kern-.025em b}\kern-.08em
    T\kern-.1667em\lower.7ex\hbox{E}\kern-.125emX}}
\begin{document}

\newcommand{\felix}[1]{{#1}}
\newcommand{\fabian}[1]{{#1}}
\newcommand{\marvin}[1]{{#1}}
\newcommand{\edited}[1]{{#1}}

\lstdefinestyle{lststyle}{%
    escapechar=@,
    columns=fullflexible
}
\lstdefinelanguage{complexity_lan}{
    keywords = {while, do, if, then, end,}
}
\definecolor{myGrey}{RGB}{140,140,140}
\lstdefinelanguage{pseudo_lan}{
    keywords = {while, do, if, then, end, for, else, return},
    comment=[l]{\#},
    commentstyle=\color{myGrey}\ttfamily
}
\lstset{
style=lststyle,
}

\title{I Like To Move It -- Computation Instead of Data in the Brain\\
\thanks{Copyright held by IEEE DOI 10.1109/IPDPS65963.2026.00028.}
}

\author{\IEEEauthorblockN{Fabian Czappa}
\IEEEauthorblockA{\textit{Department of Computer Science} \\
\textit{Technical University of Darmstadt}\\
Darmstadt, Germany \\
ORCID: 0000-0001-8422-5706}
\and
\IEEEauthorblockN{Marvin Kaster}
\IEEEauthorblockA{\textit{Department of Computer Science} \\
\textit{Technical University of Darmstadt}\\
Darmstadt, Germany \\
ORCID: 0000-0002-6468-2122}
\and
\IEEEauthorblockN{Felix Wolf}
\IEEEauthorblockA{\textit{Department of Computer Science} \\
\textit{Technical University of Darmstadt}\\
Darmstadt, Germany \\
ORCID: 0000-0001-6595-3599}
}

\maketitle

\begin{abstract}
The detailed functioning of the human brain remains incompletely understood. 
Large-scale brain simulations complement experimental research but face substantial computational challenges: the human brain comprises approximately $10^{11}$ neurons connected by $10^{14}$ synapses, collectively forming the connectome.
Empirical evidence indicates that modifications of the connectome---specifically the formation and elimination of synapses, referred to as structural plasticity---are essential for processes such as learning and memory formation. 
Connectivity updates can be computed efficiently using a Barnes--Hut-inspired approximation that reduces computational complexity from $O(n^2)$ to $O(n \log n)$, where $n$ denotes the number of neurons.
Despite this improvement, communication overhead still limits scalability. 
Synapse updates rely heavily on remote memory access (RMA), and spike transmission requires all-to-all communication at every simulation time step.
We introduce a novel algorithm that reduces communication by migrating computation rather than data. 
This approach reduces connectivity update time by a factor of 6 and spike transmission time by more than 2 orders of magnitude.
\end{abstract}

\section{Introduction}
\label{sec:introduction}
Despite being directly involved in our daily activities, comprehending the human brain remains challenging because of its estimated 86 billion neurons, 100 trillion synapses (neural connections), and an uncertain count of non-neuronal cells~\cite{herculano2009human}. Its internal mechanisms and various functions are still largely mysterious, including the exact neuron count~\cite{10.1093/brain/awae390}. Nevertheless, some elements are established. For instance, there is compelling evidence that synapses adapt to injuries~\cite{Butz2013}, learning~\cite{bennett2018rewiring, la2020brain}, and developmental stages~\cite{FANDAKOVA2020100764}. Given this reality, combined with ethical issues and experimental constraints involving human participants, modeling brain development through simulations becomes crucial for grasping the broader organizational principles at work.

This creates an additional difficulty: While many models permitting alterations in brain connectivity (the \textit{structural connectome}) maintain fixed synaptic arrangements and modify synaptic weights (referred to as \textit{synaptic plasticity})~\cite{citri2008synaptic, magee2020synaptic}, models describing connectivity changes are uncommon~\cite{BUTZ2009287}. 
Some concentrate on individual neuronal growth (overlooking the bigger picture)~\cite{tavosanis2012dendritic, yamada2016structural}, whereas others treat neurons as basic points and emphasize overall dynamics~\cite{10.1371/journal.pcbi.1009836, 10.3389/fninf.2024.1323203}. 
\edited{A simulation that mimics all cells in the human brain and their interactions is far from achievable with our current understanding and the limits of current hardware.}
A notable example of the latter class, which simplifies the neurons to focus on their overall dynamics, is the Model of Structural Plasticity (MSP)~\cite{Butz2013}, which is distinctive in its area: 
It assumes neurons possess target activities and adjust their connectivity according to their present and desired activity levels. When neuronal activity falls short of the target, they form new synapses to increase input; conversely, when activity exceeds the target, they eliminate synapses to decrease input. 
This homeostatic principle leads the entire brain to equilibrium with minimal modification, where most neurons operate near their individual targets. 
Yet this approach generates computational limitations from a computer science viewpoint, since parts of the model exhibit quadratic time complexity (relative to neuron count): 
Synapse formation requires computing a probability for each potential connection from the originating neuron. 
Prior research implements the Barnes--Hut algorithm~\cite{barnes1986hierarchical, Rinke2018} or the fast multipole method~\cite{darve2000fast, noettgen_ea:2022:relearn-fmm} to decrease computational complexity to nearly linear~\cite{CZAPPA202324}. 
Moreover, upon neuron spiking, notification must reach all connected neurons, as the spike travels across all outgoing synapses. 
Theoretically, this involves a time complexity equal to the number of neurons multiplied by synapses per neuron; practically, however, all processes that partake in the simulation still need to interact with one another.

In this study, we introduce two algorithms designed to enhance MSP; one extends the Barnes--Hut algorithm used in prior research~\cite{Rinke2018}, and another approximates neuronal spikes. 
Our approach, like nearly all preceding work, employs the Message Passing Interface (MPI)~\cite{mpi41} as a standard framework for process communication, workload distribution, and data sharing. 
Initially, we enhance the Barnes--Hut algorithm to be \textit{location aware} as follows: 
Previously, neurons were organized within a spatial octree, which was itself distributed~\cite{Rinke2018}. 
The upper section was duplicated across all MPI ranks, while subtrees beneath a specific level were allocated to different MPI ranks. 
When a neuron required data from an MPI rank other than the one managing the neuron, it retrieved the data through \fabian{remote memory access (RMA)}. 
\fabian{This allows the requesting MPI rank to access the data without active involvement of the ``sending'' MPI rank, simplifying the communication scheme.}
In our improved version, we transfer the neuron to the appropriate MPI rank and relocate the computation rather than moving the data. 
While this does not decrease the computational complexity, it reduces the communication complexity from logarithmic dependence on the neuron count to constant per neuron.

Secondly, we reduce the bottleneck caused by synchronizing the spiking behavior every time step.
Instead, we propose an algorithm to approximate the firing rates of neurons by their frequency.
This allows us to drastically reduce the number of synchronization points while preserving the general activity pattern.
In contrast to our new connectivity update algorithm, this one alters the simulation results concerning the underlying model, as it introduces an additional layer of approximation.
While, in theory, this could introduce ripple effects that drastically change the dynamics of the network, in practice, this is not the case.
We observe a slight deviation between our approximation and the ``ground truth'' (also only a model), but this difference has no wider practical consequences.

In Section~\ref{sec:related}, we review work on neuron models, other simulations of plasticity, full-scale simulations of neuronal activity, and applications of the Barnes--Hut algorithm in computational biology.
Following this, in Section~\ref{sec:background}, we introduce all relevant models for our work as background, including basic terminology.
In Section~\ref{sec:implementation}, we introduce our two algorithms, explain their pros and cons, and dive into implementation details.
Our evaluation in Section~\ref{sec:results} focuses on the speed of our algorithms compared to the status quo, the qualitative changes of the simulation results, and a theoretical analysis of our optimizations.
Lastly, we conclude our work in Section~\ref{sec:conclusion} and give an outlook on what can and should be done (on the computer-science side) in the future of whole-brain structural plasticity.

\section{Related Work}
\label{sec:related}
There exists a wide variety of brain simulators tailored to different use cases and research questions.
NEST~\cite{Gewaltig:NEST} provides a general-purpose framework for simulating spiking neurons; however, in most scenarios, neuronal connectivity remains static.
Although it permits manual connectivity modifications, it also incorporates MSP~\cite{Butz2013} with a time complexity of $O(n^2)$ where $n$ represents the number of neurons (i.e., without the Barnes--Hut algorithm), which restricts scalability severly.
Other brain simulators, like Arbor~\cite{paper:arbor2019}, permit connectivity modifications and model neurons with considerably greater detail.
For Arbor specifically, one must program the code to manage plasticity manually.
Therefore, theoretically, any model is implementable.
NEURON~\cite{carnevale2006neuron} emphasizes morphologically detailed models, encompassing multi-compartment models, ion channels, and cable equations.
Brian2~\cite{stimberg2019brian}, conversely, is designed for simple neuron morphology but allows users to define differential equations themselves, providing high flexibility and ease of extension.
CARLSim~\cite{niedermeier2022carlsim} is a simulator engineered for large-scale simulations on heterogeneous clusters supporting both CPU and GPU architectures.
Among these simulators, only NEST natively supports structural plasticity.
The remaining simulators enable users to modify connectivity, but require users to code the logic determining when and where synapses should be formed or eliminated.
Numerous additional simulators exist that do not simulate individual neurons, such as the Virtual Brain~\cite{10.3389/fninf.2013.00010}, which targets full-brain dynamics at much reduced resolution (and without connectivity modifications), or those that do not model individual spikes, like Nengo~\cite{bekolay2014nengo}.

Other large-scale simulations encompass a 2009 model featuring $10^{9}$ neurons with $10^{13}$ synapses~\cite{6375547}, a 2014 NEST simulation with $1.86 \cdot 10^{9}$ neurons and  $1.11 \cdot 10^{13}$ synapses~\cite{10.3389/fninf.2014.00078}, and a model with $6.8 \cdot 10^{10}$ neurons and $5.4 \cdot 10^{13}$ synapses~\cite{10.3389/fninf.2020.00016}.
Nevertheless, all these large-scale simulations maintain fixed neuronal connectivity, eliminating the primary computational burden from the simulation—namely, flexible spike communication structures and general methods for arbitrary neuronal connections during simulation.

MSP exhibits an \textit{n-body problem} structure: given $n$ bodies, compute their interactions.
This structure is common in physics~\cite{Pfalzner_Gibbon_1996}, where interacting entities range from elementary particles to solar systems and beyond~\cite{COSTA200517}.
Typically, the solution carries a computational time complexity of $O(n^{2})$, since each body interacts with every other body.
Numerous approximation approaches have already been investigated.
For instance, acting bodies can be clustered if sufficiently distant—over large distances, it matters little whether a body is slightly mispositioned or whether a single force (double the potential) replaces two separate forces.
This concept forms the foundation of the Barnes--Hut algorithm~\cite{barnes1986hierarchical, barnes1990modified, burtscher2011efficient}, which aggregates bodies and constructs a spatial octree (root represents entire domain, eight children represent octants, etc.).
The algorithm utilizes a parameter $\theta$ (typically $[0.1, 0.4]$) and approximates received forces from a subdomain if sufficiently small—the ratio of subdomain length to distance from center of mass must be less than $\theta$.
With $\theta > 0$, the Barnes--Hut algorithm achieves $O(n \cdot \log{n})$ time complexity, while $\theta = 0$ yields the direct solution.
Clustering receiving bodies leads to the fast multipole method (FMM)~\cite{darve2000fast, ROKHLIN1985187}, operating with $O(n)$ time complexity.
Shifting perspective from an algorithmic to an absolute standpoint, the fundamental objective is to (implicitly) compute the $n \times n$ interaction matrix.
Should this matrix be low-rank, numerical linear algebra techniques can be applied for efficient computation~\cite{doi:10.1137/23M1565103}, though this condition does not hold for our problem.

MSP presents a unique challenge that physics problems typically avoid: whereas for a star, the gravitational force exerted by a distant galaxy suffices, MSP requires more specific information.
It is insufficient to merely know that a new synapse extends into a brain region---the new partner must be a genuine neuron.
Consequently, prior research has modified the Barnes--Hut algorithm~\cite{Rinke2018}, applying it recursively until an actual neuron is identified (which maintains the time complexity of $O(n \cdot \log{n})$~\cite{CZAPPA202324}).
FMM can also be adjusted~\cite{noettgen_ea:2022:relearn-fmm}; however, due to the necessity of finding an actual partner, the modified version likewise exhibits a time complexity of $O(n \cdot \log{n})$.
Additionally, because neurons require actual targets, FMM limits the selection of closely positioned neurons searching for partners simultaneously—they must connect to the same region.

The Barnes--Hut algorithm finds application in various areas of computational biology beyond modeling structural plasticity in the brain.
For instance, early studies employed it for approximating molecular dynamics~\cite{PLIMPTON19951}.
It can also be applied outside its primary domain, such as accelerating t-SNE classification~\cite{RUKHSAR2023104833} or visualization~\cite{hadjiabadi2021maximally} to aid in seizure or epilepsy models.

\section{Background}
\label{sec:background}
We recapitulate definitions, observations, and algorithms from publications such as the one introducing the MSP~\cite{Butz2013} and its approximation~\cite{Rinke2018}.
We will focus on the source of parallelism when using the Barnes--Hut algorithm, and review how a spatial octree is used to speed up the calculation.

\subsection{The model of structural plasticity}
\label{ssec:msp}
The model does what its name promises, i.e., it implements a model of structural plasticity: 
How neurons form new synapses and delete existing synapses over a fixed period.
It comprises three distinct internal phases that cycle until completion:
\begin{enumerate}
    \item The update of electrical activity.
    \item The update of synaptic elements.
    \item The update of connectivity.
\end{enumerate}
The update of the synaptic elements affects both a neuron's axon (its part that sends out signals to other neurons) and its dendrite (i.e., the part that receives signals from other neurons).
However, the update of synapses only happens every $100^{\text{th}}$ time step as a measure against high fluctuations.

\paragraph{Update of electrical activity}
\label{par:electricalactivity}
First, neurons exchange spikes from the previous update step (computed using models like Izhikevich~\cite{izhikevich}). 
These spikes travel across synapses to connected neurons through their axons.
Next, each neuron calculates its new electrical state using its current electrical state and incoming signals (through dendrites or as background noise).
If the neuron model detects that a neuron has spiked, this spike is recorded for the next time step, when it is transmitted via the neuron's axon.
Finally, neurons update their intracellular calcium levels—a running average of firing rates that serves as a dampening mechanism.

\paragraph{Update of synaptic elements}
\label{par:synapticlements}
Each neuron updates the synaptic elements on its axon and dendrite.
This update depends on the current and target calcium levels. 
A calcium level below the target promotes growth of synaptic elements, while a level above the target causes retraction of synaptic elements.

\paragraph{Update of synapses}
\label{par:synapses}
This phase consists of two sub-phases.
First, if a neuron retracts a vacant synaptic element, it is simply removed without affecting any synapses. 
However, if a neuron retracts a synaptic element while all are bound in synapses, one is chosen randomly, the synapse is broken, and the affected partner gains a new vacant synaptic element (and loses one synapse).
Second, all neurons with vacant axonal elements search for matching vacant dendritic elements on other neurons. 
This is done by evaluating a probability kernel based on the distance between the searching neuron and potential partners; one potential partner is selected randomly according to the probability distribution, and a synapse is proposed.
Once all synapses are proposed, neurons accept as many new synapse requests as possible (randomly) based on their vacant dendritic elements and the number of received requests. 
Requests that cannot be fulfilled because a neuron received more requests than it has vacant dendritic elements are declined. 
The axonal elements receive the response (success/failure) and either form a synapse or remain vacant for the next synaptic update.

\subsection{The Barnes--Hut algorithm}
\label{ssec:barneshut}%
\begin{figure}
\begin{tikzpicture}
\draw[black, thick] (3,1) rectangle (3.5,1.5) node[pos=.5] {1};

\draw[->, black] (3.25,1) -- (1.25,0.5);
\draw[->, black] (3.25,1) -- (5.25,0.5);

\draw[black, thick] (1,0) rectangle (1.5,0.5) node[pos=.5] {1};
\draw[black, thick] (5,0) rectangle (5.5,0.5) node[pos=.5] {2};

\draw[->, black] (1.25,0) -- (0.25,-0.5);
\draw[->, black] (1.25,0) -- (2.25,-0.5);
\draw[->, black] (5.25,0) -- (4.25,-0.5);
\draw[->, black] (5.25,0) -- (6.25,-0.5);

\draw[black, thick] (0,-1) rectangle (0.5,-0.5) node[pos=.5] {1};
\draw[black, thick] (2,-1) rectangle (2.5,-0.5) node[pos=.5] {1};
\draw[black, dashed] (4,-1) rectangle (4.5,-0.5) node[pos=.5] {2};
\draw[black, dashed] (6,-1) rectangle (6.5,-0.5) node[pos=.5] {2};

\draw[->, black] (0.25,-1) -- (-0.35,-1.5);
\draw[->, black] (0.25,-1) -- (0.85,-1.5);
\draw[->, black] (2.25,-1) -- (1.65,-1.5);
\draw[->, black] (2.25,-1) -- (2.85,-1.5);
\draw[->, black, dashed] (4.25,-1) -- (3.65,-1.5);
\draw[->, black, dashed] (4.25,-1) -- (4.85,-1.5);
\draw[->, black, dashed] (6.25,-1) -- (5.65,-1.5);
\draw[->, black, dashed] (6.25,-1) -- (6.85,-1.5);

\draw[black, thick] (-0.6,-1.5) rectangle (-0.1,-2) node[pos=.5] {1};
\draw[black, thick] (0.85,-1.75) circle (0.25) node {1};
\draw[black, thick] (1.65,-1.75) circle (0.25) node {1};
\draw[black, thick] (2.6,-1.5) rectangle (3.1,-2) node[pos=.5] {1};
\draw[black, dashed] (3.65,-1.75) circle (0.25) node {2};
\draw[black, dashed] (4.8,-1.75) circle (0.25) node {2};
\draw[black, dashed] (5.65,-1.75) circle (0.25) node {2};
\draw[black, dashed] (6.6,-1.5) rectangle (7.1,-2) node[pos=.5] {2};

\draw[->, black] (-0.35,-2) -- (-0.35,-2.5);
\draw[->, black] (-0.35,-2) -- (0.85,-2.5);
\draw[->, black] (2.85,-2) -- (2.85,-2.5);
\draw[->, black, dashed] (6.85,-2) -- (5.65,-2.5);
\draw[->, black, dashed] (6.85,-2) -- (6.85,-2.5);

\draw[black, thick] (-0.35,-2.75) circle (0.25) node {1};
\draw[black, thick] (0.85,-2.75) circle (0.25) node {1};
\draw[black, thick] (2.85,-2.75) circle (0.25) node {1};
\draw[black, dashed] (5.65,-2.75) circle (0.25) node {2};
\draw[black, dashed] (6.85,-2.75) circle (0.25) node {2};

\node[rotate=90, text width=1cm] at (-1,0.4) {branch nodes};

\end{tikzpicture}
\caption{Example of a distributed tree (shown is a binary tree for the sake of simplicity) from the view of the first process. Boxes are inner nodes, and circles are leaf nodes; the numeric label indicates the process of owning the node. As we depict the view from process 1, it owns the (replica of the) root node. Solid lines indicate information on process 1, and dashed lines indicate information that resides on other ranks.}
\label{fig:tree:distributed}
\end{figure}
The most time-consuming aspect of MSP is the synaptic update, even though it only occurs every 100 steps. 
Since each vacant axonal element must calculate its probability of connecting to every vacant dendritic element, this can result in quadratic time complexity relative to the number of neurons.
To address this, previous work adapted the Barnes--Hut~\cite{Rinke2018} algorithm to reduce computational complexity to quasi-linear~\cite{CZAPPA202324}. 
This is achieved through (a) dividing the simulation domain, (b) creating a spatial octree, and (c) distributing the computational work.

\paragraph{Division of the simulation domain}
\label{par:simulationdomain}
Given a number of MPI ranks $k$ (a power of two) and  a simulation domain with bounds $D = [0, l_x] \times [0, l_y] \times [0, l_z] \subseteq \mathbb{R}^3$, find the smallest $b$ such that $8^{\,b-1} \leq k < 8^{\,b}$, and divide $D$ into $8^{\,b}$ subdomains, each of size $(l_x / b) \times (l_y / b) \times (l_z /b)$, which then cover $D$.
These subdomains are indexed by a space-filling curve (e.g., the Morton curve), and each MPI rank is responsible for 1, 2, or 4 consecutive subdomains (in the sense of the space-filling curve).

\paragraph{Spatial octree}
\label{par:spatialoctree}
The MPI ranks jointly build a spatial octree covering all neurons in the simulation domain.
They share a common upper portion:
The root node represents the entire simulation domain; its eight children represent the eight octants created by splitting the simulation domain along the x-, y-, and z-axes, continuing this process until level $b$ (as determined previously).
Level $b$ contains the \textit{branch nodes}, each representing one of the $8^{\,b}$ subdomains.
Moving deeper into the tree from level $b+1$, only the MPI rank responsible for a subdomain holds actual data.
That rank subdivides its subdomain until each cell (leaf node in the octree) contains at most one neuron.
Figure~\ref{fig:tree:distributed} illustrates such a distributed tree.

\paragraph{Division of work}
\label{par:divisionwork}
Before the synaptic update, the MPI ranks update the distributed octree based on vacant synaptic elements. 
Leaf nodes without actual neurons have no valid position and no vacant dendritic elements; those with neurons have the neuron's position as their own and as many vacant dendritic elements as the neuron. 
Inner nodes contain as many vacant dendritic elements as all their children combined, with their position being the weighted average of their children. 
The MPI ranks update their octree from bottom-up to the branch nodes, where they perform all-to-all exchanges of branch nodes and then continue updating up to the root node. 
This approach ensures each MPI rank knows how many vacant synaptic elements exist up to the branch node level and approximately where they are located (based on average positions).

When a neuron has a vacant axonal element and wants to form a synapse, the MPI rank containing that neuron handles all resulting calculations.
It begins with the root node as a potential target, applies the Barnes--Hut algorithm's acceptance criterion (cell length divided by distance must be smaller than $\theta$; this always rejects the root node), and recursively replaces a node's eight children when the node is rejected.
One child is randomly selected based on connection probability from a list of nodes meeting the acceptance criterion.
Everything proceeds smoothly if the target is a leaf node and a synapse-forming request is generated.
This request includes:
\begin{itemize}
    \item The source neuron's ID
    \item The target neuron's ID
    \item The target node's cell type (excitatory or inhibitory)
\end{itemize}
However, if the target node is an inner node, the entire process restarts with the target node instead of the root node.
When the MPI rank needs octree nodes it does not own, it downloads them via RMA and attaches them to its own octree; these remain valid until the end of the synapse-formation phase and thus do not need re-downloading for subsequent neurons requiring them.

Once all formation requests have been formed, these are all-to-all submitted, locally accepted/declined, and their responses are returned.
Note here that a simple \textit{yes}/\textit{no} is sufficient as an answer, as the requesting neuron knows which partner it has chosen.

\section{Communication-Efficient Algorithms}
\label{sec:implementation}
In this section, we explain the reasoning behind both proposed algorithms for updating connectivity and exchanging neuron spikes. 
We will demonstrate our reduction in (theoretical) communication complexity and (practical) communication implementation.

\subsection{The location-aware Barnes--Hut algorithm}
\label{ssec:locationaware}
\begin{figure}
\begin{tikzpicture}
\draw[black, thick] (3,1) rectangle (3.5,1.5) node[pos=.5] {1};

\draw[->, black] (3.25,1) -- (1.25,0.5);
\draw[->, black] (3.25,1) -- (5.25,0.5);

\draw[black, thick] (1,0) rectangle (1.5,0.5) node[pos=.5] {1};
\draw[black, thick] (5,0) rectangle (5.5,0.5) node[pos=.5] {2};

\draw[->, black] (1.25,0) -- (0.25,-0.5);
\draw[->, black] (1.25,0) -- (2.25,-0.5);
\draw[->, black] (5.25,0) -- (4.25,-0.5);
\draw[->, black] (5.25,0) -- (6.25,-0.5);

\draw[black, thick] (0,-1) rectangle (0.5,-0.5) node[pos=.5] {1};
\draw[black, thick] (2,-1) rectangle (2.5,-0.5) node[pos=.5] {1};
\draw[black, dashed] (4,-1) rectangle (4.5,-0.5) node[pos=.5] {2};
\draw[black, dashed] (6,-1) rectangle (6.5,-0.5) node[pos=.5] {2};

\draw[->, black] (0.25,-1) -- (-0.35,-1.5);
\draw[->, black] (0.25,-1) -- (0.85,-1.5);
\draw[->, black] (2.25,-1) -- (1.65,-1.5);
\draw[->, black] (2.25,-1) -- (2.85,-1.5);
\draw[->, black, dashed] (4.25,-1) -- (3.65,-1.5);
\draw[->, black, dashed] (4.25,-1) -- (4.85,-1.5);
\draw[->, black, dashed] (6.25,-1) -- (5.65,-1.5);
\draw[->, black, dashed] (6.25,-1) -- (6.85,-1.5);

\draw[black, thick] (-0.6,-1.5) rectangle (-0.1,-2) node[pos=.5] {1};
\draw[black, thick] (0.85,-1.75) circle (0.25) node {1};
\draw[black, thick] (1.65,-1.75) circle (0.25) node {1};
\draw[black, thick] (2.6,-1.5) rectangle (3.1,-2) node[pos=.5] {1};
\draw[black, dashed] (3.65,-1.75) circle (0.25) node {2};
\draw[black, dashed] (4.8,-1.75) circle (0.25) node {2};
\draw[black, dashed] (5.65,-1.75) circle (0.25) node {2};
\draw[black, dashed] (6.6,-1.5) rectangle (7.1,-2) node[pos=.5] {2};

\draw[->, black] (-0.35,-2) -- (-0.35,-2.5);
\draw[->, black] (-0.35,-2) -- (0.85,-2.5);
\draw[->, black] (2.85,-2) -- (2.85,-2.5);
\draw[->, black, dashed] (6.85,-2) -- (5.65,-2.5);
\draw[->, black, dashed] (6.85,-2) -- (6.85,-2.5);

\draw[black, thick] (-0.35,-2.75) circle (0.25) node {1};
\draw[black, thick] (0.85,-2.75) circle (0.25) node {1};
\draw[black, thick] (2.85,-2.75) circle (0.25) node {1};
\draw[black, dashed] (5.65,-2.75) circle (0.25) node {2};
\draw[black, dashed] (6.85,-2.75) circle (0.25) node {2};

\node[rotate=90, text width=1cm] at (-1,0.4) {branch nodes};

\draw[blue, thick] (-0.35,-2.75) circle (0.35);
\draw[red, thick, dotted] (3.9,-1.1) rectangle (4.6,-0.4);
\draw[red, thick, dotted] (5.9,-1.1) rectangle (6.6,-0.4);
\draw[red, thick, dotted] (5.65,-1.75) circle (0.35);
\draw[red, thick, dotted] (6.5,-1.4) rectangle (7.2,-2.1);
\draw[red, thick, dotted] (5.65,-2.75) circle (0.35);
\draw[red, thick] (6.85,-2.75) circle (0.35);

\end{tikzpicture}
\caption{The neuron from process 1 (the leaf node with the blue solid marking) will propose a synapse to the neuron from process 2 (the leaf node with the red solid marking). In the old algorithm, process~1 must download all red nodes via remote memory access. In our proposed algorithm, process~1 sends only a part of the information from the neuron to process~2. Thus, the transfer direction has mostly been reversed, and the computation has been moved to the target node.}
\label{fig:tree:download}
\end{figure}
\begin{algorithm} 
\lstset{basicstyle=\ttfamily\footnotesize,breaklines=true}
\begin{lstlisting}[numbers=left, stepnumber=1, language= pseudo_lan]
update_octree(max, branch_level) # from - to
all_exchange_branch_nodes()
update_octree(branch_level, 0)   # from - to

intermediate_results = []

foreach local neuron n:
    # terminate search early 
    pos = n.position
    t = barnes-hut(root, branch_level, pos)
    intermediate_results += t, n.position

# send intermediate_results to ranks owning t
local_requests = all_exchange(intermediate_results)
local_responses = []

foreach local_root, pos in local_requests:
    r = barnes-hut(local_root, max, pos)
    local_responses += r

# send results back to ranks 
results = all_exchange(local_responses)

foreach r in results:
    if r.successful:
        add_synapse(r.source, r.target)
\end{lstlisting}
\caption{Pseudocode for the main algorithm, including the location-aware aspect of sending computation requests.}
\label{alg:bhla} 
\end{algorithm} 
The primary bottleneck in the work division~(\ref{par:divisionwork}) is not the actual computations but the waiting and data transfer times.
Although RMA eliminates the need for the ``sender'' to manage communication, data transfer still must occur.
Furthermore, the algorithm's structure does not facilitate hiding waiting times through other computations:
When requesting \textit{far} nodes, computation has already progressed to a point where it can only continue once the data arrives.

In our work, we propose a hybrid algorithm that significantly reduces communication:
The first step remains identical to~\ref{par:divisionwork}:
A neuron with a vacant axonal bouton searches for a partner starting at the root node.
If the returned node is above or below the branch node level, this process repeats until a target node at or below the branch level is found.
If nodes stored on other MPI ranks are needed during these searches, we download them with MPI as before.
However, if the returned node is at or below the branch node level, we instead send a request for \textit{synapse formation and calculation} to the MPI rank responsible for that node.
This request contains:
\begin{itemize}
    \item The ID of the source neuron
    \item The position of the source neuron
    \item The ID of the target node
    \item The type of target node; i.e., if the target is already a leaf
    \item The cell type of the target node; i.e., excitatory or inhibitory
\end{itemize}
All MPI ranks construct such requests and all-to-all exchange them.
In the implementation, an old request has a size of \texttt{8 + 8 + 1 = 17}~B, while the new request has a size of \texttt{8 + 24 + 8 + 1 + 1 = 42}~B. 

Upon receiving such requests, the MPI ranks check the target node type.
If it is a leaf, they need not compute anything and can convert the request to the old format (source ID, target ID, type).
Otherwise, the target node is an inner node of the octree (note that we are on the MPI rank owning that tree portion).
The MPI rank owning the target node begins searching at the specified node using the node's position that wants to form a synapse (i.e., from the other MPI rank).
This search requires no additional RMA communication but involves a more extensive response:
Instead of \textit{yes}/\textit{no} (\texttt{1}~B), the response now includes the ID of the found neuron (if any) and an indicator of search success (as before, too many requests may reach a neuron, causing it to decline some)—totaling \texttt{8 + 1 = 9}~B.
Figure~\ref{fig:tree:download} illustrates a schematic comparison of communication requirements between the old and new algorithms, \fabian{while Algorithm~\ref{alg:bhla} shows the pseudocode of the new method.}
\subsection{Approximation of firing rates}
\label{ssec:firingrate}
Beyond synapse formation, the second major bottleneck in the simulation is spike exchange.
When a neuron spikes, the action potential travels through its axon across synapses to other neurons' dendrites, which then receive the spike as input in the next simulation step.
Strictly speaking, this is not all-to-all communication, as neurons have a fixed number of partners regardless of total neuron count. 
However, in practice, this becomes all-to-all communication among involved MPI ranks because neurons typically maintain connections to thousands of other neurons, requiring communication with most MPI ranks.

Any standard neuron model is \textit{time-critical} concerning the received spiking times, i.e., the state transitions $s_0 \xrightarrow[\text{spike}]{\text{receives}} s_1 \xrightarrow[\text{no spike}]{\text{receives}} s_2$ and $s_0 \xrightarrow[\text{no spike}]{\text{receives}} s_1' \xrightarrow[\text{spike}]{\text{receives}} s_2'$ result in different ending states.
Only for the most fundamental state transitions would $s_2 = s_2'$ hold (e.g., simple combinations of minimum, sum, maximum, ...), although even in such cases, $s_1$ and $s_1'$ need not be equal.
In general, state transitions with different numbers of spikes result in different ending states as well.

However, the previous model for neuronal spikes already approximates various aspects.
In reality, spikes are not transmitted instantaneously; they occur over extended periods, and their effects depend on spike travel distance, exact contact positions on receiving neurons, and other factors.
\edited{Furthermore, real synapses can be electrical or chemical, and glial cells also influence the firing activity.}
Keeping this in mind, we propose approximating neuron model spikes using firing frequency—instead of individual spikes, frequencies are transmitted periodically, and receiving neurons use pseudo-random number generators and these frequencies to determine if the sending neuron spiked.
We only apply this model for spike transmission between MPI ranks, as checking whether one spiked is virtually free for connected neuron pairs on the same MPI rank.

This algorithm significantly reduces the number of synchronization points.
Instead of synchronizing and exchanging data at every time step, we define an epoch length $\Delta$ and exchange neuron firing frequencies only every $\Delta$ time steps.
This approach has some drawbacks, though none substantially impact the simulation.
First, exchanging frequencies might involve more data transfer than $\Delta$ spike transmissions (due to different data types and necessity to send data between each connected neuron pair rather than only those where the sending neuron spiked).
Second, since neurons do not maintain steady firing frequencies during the development phase, our model introduces response lag.
Lastly, neuron spikes are no longer synchronized.
For example, given a neuron that spikes every ten time steps and three connected neurons (one on the same MPI rank and two on different MPI ranks), the first receives actual spikes while the other two determine at each time step with ten percent probability whether the neuron spiked.
Theoretically, the received spikes need not correlate and could have arbitrarily large mismatches—though they do correlate in practice.

\section{Results}
\label{sec:results}
We will start our results with theoretical observations regarding the algorithm complexity changes. 
Then, we will examine our timing results on actual hardware, and finally, we will briefly analyze the amount of transferred data.
\fabian{For the timings, we provide \emph{weak-scaling} results (keeping the number of neurons fixed per MPI rank, and increasing the number of MPI ranks) and \emph{strong-scaling} results (keeping the total number of neurons fixed, and increasing the number of MPI ranks}.
\begin{figure*}
\centering
\includegraphics[width=\linewidth]{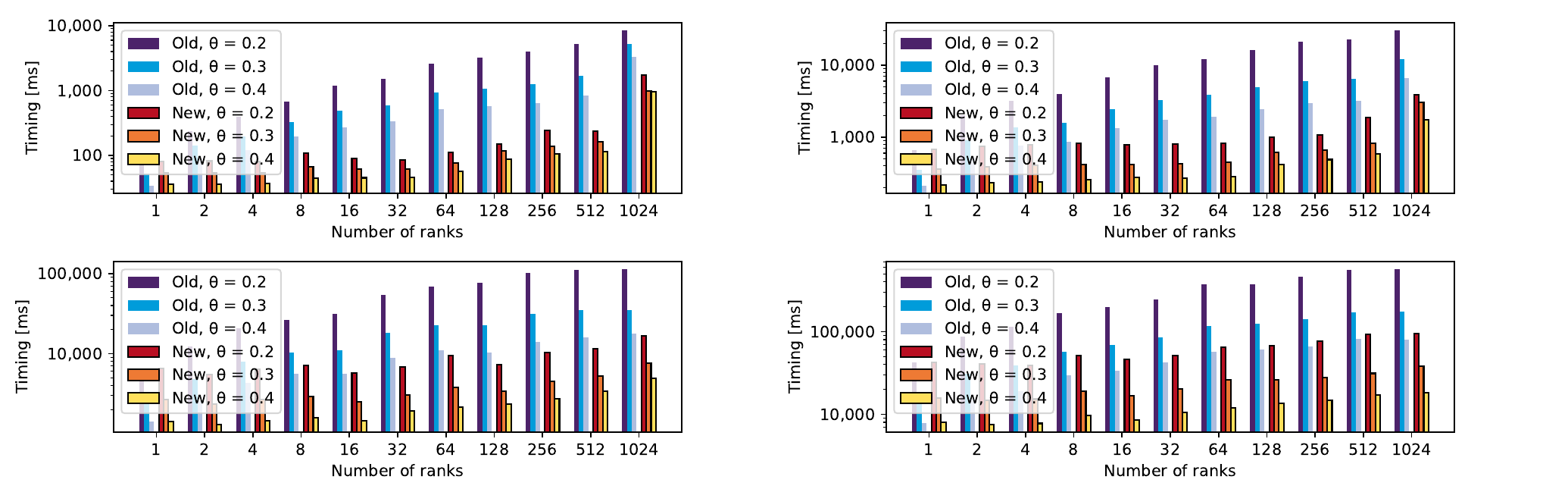}
\caption{Timing plots for the \textit{old} Barnes--Hut algorithm and our proposed \textit{new} location-aware Barnes--Hut algorithm, given for varying numbers of neurons per rank: 1024 neurons per MPI rank on the top left, 4096 neurons per rank on the top right, 16~384 neurons per rank on the bottom left, and 65~536 neurons per rank on the bottom right.}
\label{fig:timings:bh}
\end{figure*}
\begin{figure*}
\centering
\includegraphics[width=\linewidth]{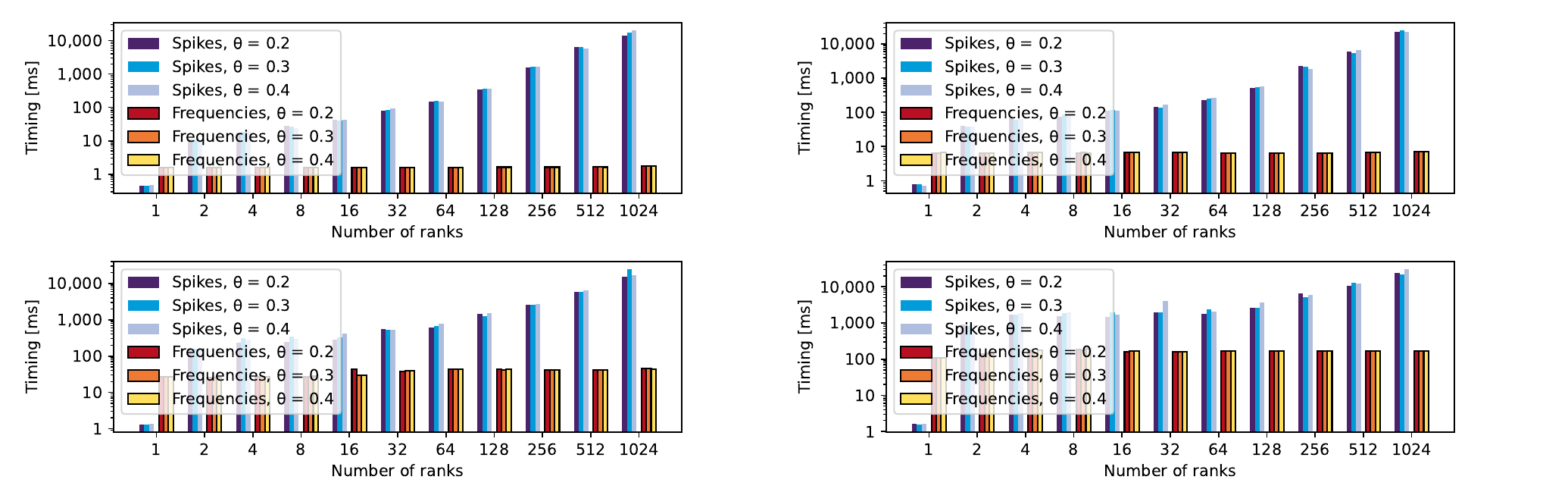}
\caption{Timing plots for the transfer of neuron \textit{spikes} and neuron firing \textit{frequencies}, given for varying numbers of neurons per MPI rank: 1024 neurons per rank on the top left, 4096 neurons per rank on the top right, 16~384 neurons per rank on the bottom left, and 65~536 neurons per rank on the bottom right. Note that transferring the frequencies is virtually free.}
\label{fig:timings:firing}
\end{figure*}
\begin{figure*}
\centering
\includegraphics[width=\linewidth]{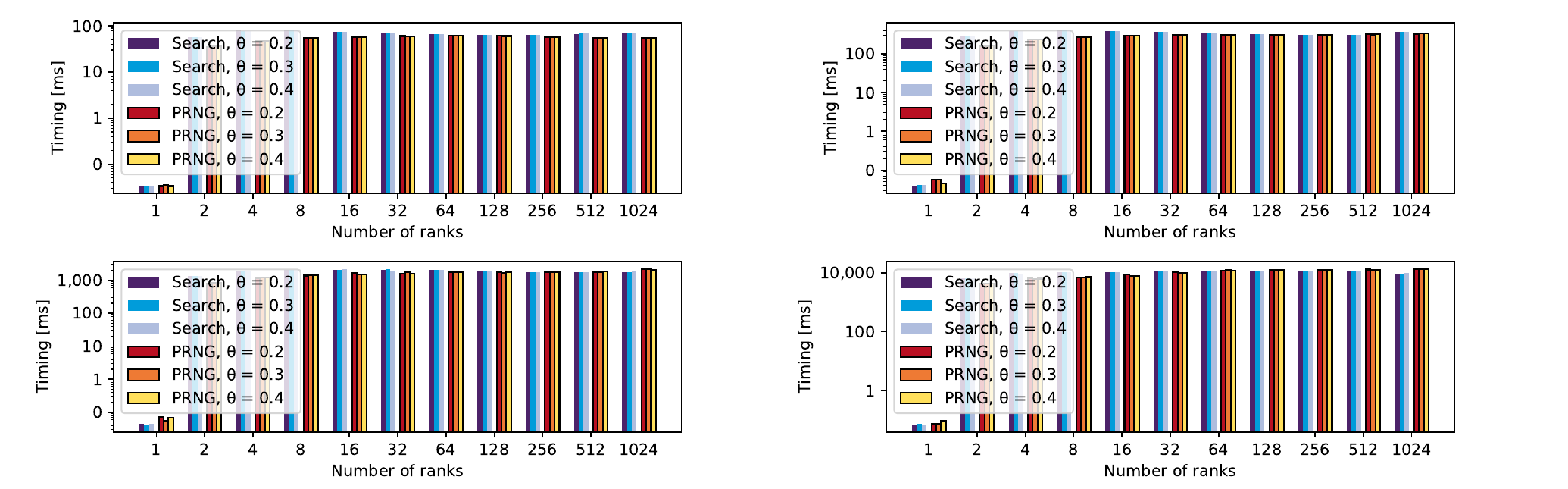}
\caption{Timing plots for look-up of spikes from distant neurons (by a binary \textit{search}) and for the approximation of spikes based on the frequency with a pseudo-random number generator (\textit{PRNG}), given for varying numbers of neurons per MPI rank: 1024 neurons per rank on the top left, 4096 neurons per rank on the top right, 16,384 neurons per rank on the bottom left, and 65,536 neurons per rank on the bottom right.}
\label{fig:timings:distant}
\end{figure*}
\begin{figure}
    \centering
    \includegraphics[width=1\linewidth]{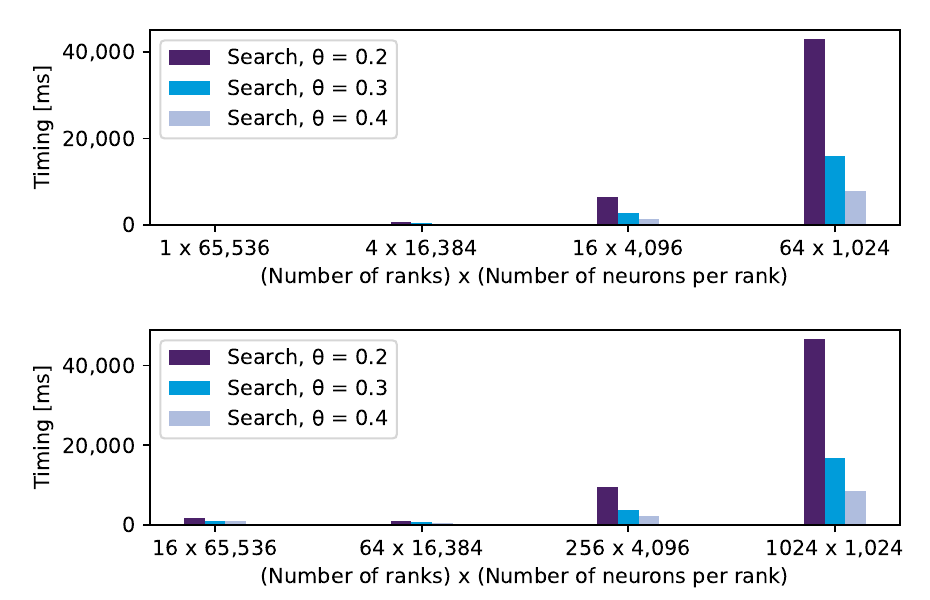}
    \caption{Timing plots for the new location-aware Barnes--Hut algorithm, with a constant number of total neurons (top: 65,536; bottom: 1,048,576). }
    \label{fig:strong:bh}
\end{figure}
\begin{figure}
    \centering
    \includegraphics[width=1\linewidth]{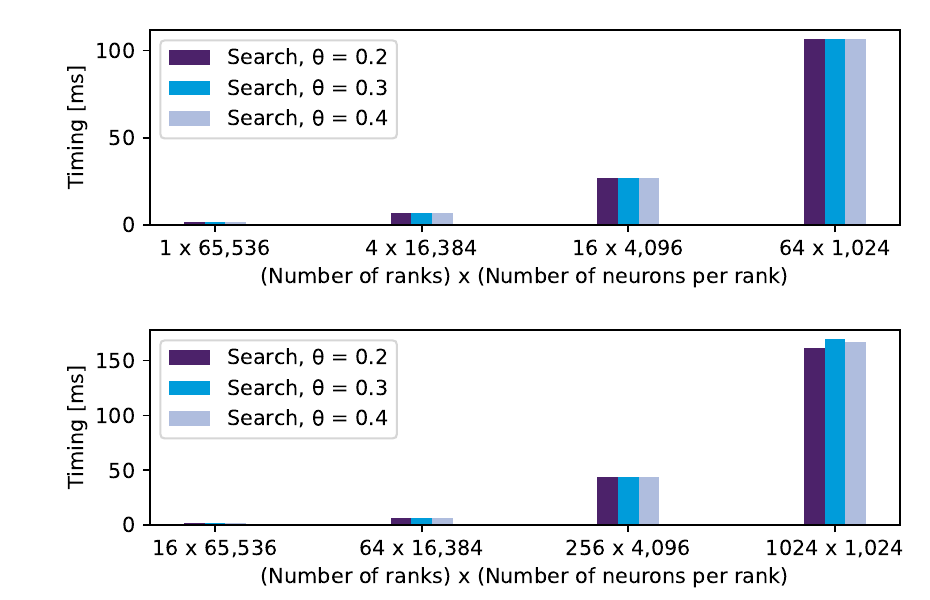}
    \caption{Timing plots for the new algorithm for transferring firing frequencies, with a constant number of total neurons (top: 65,536; bottom: 1,048,576). }
    \label{fig:strong:freq}
\end{figure}
\begin{figure}
    \centering
    \includegraphics[width=1\linewidth]{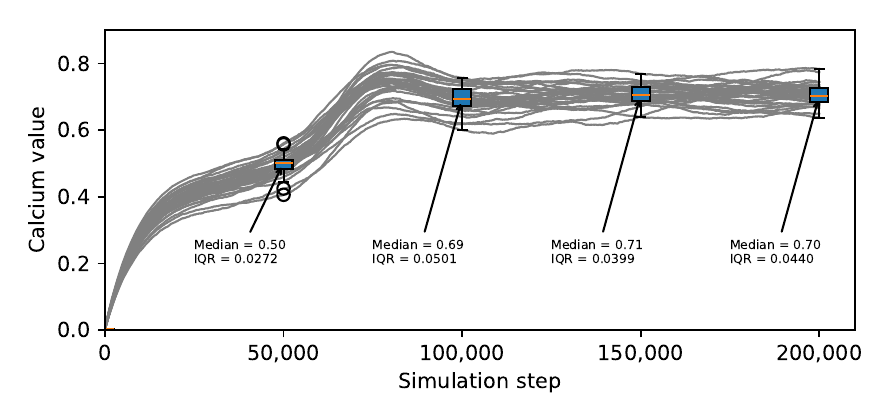}
    \caption{Calcium concentration for the 32 neurons during a simulation of 200,000 simulation steps (2000 connectivity updates). Shown is the old way of sending the firing frequencies, together with the box plots of the range every 50,000 steps.}
    \label{fig:error:old}
\end{figure}
\begin{figure}
    \centering
    \includegraphics[width=1\linewidth]{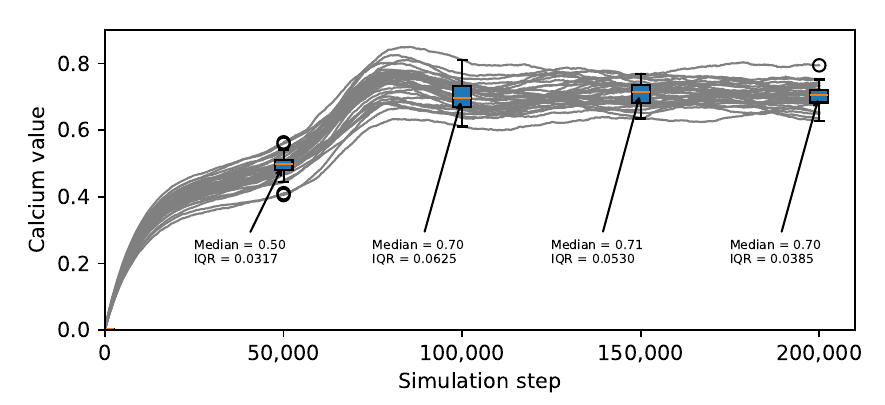}
    \caption{Calcium concentration for the 32 neurons during a simulation of 200,000 simulation steps (2000 connectivity updates). Shown is the proposed way of sending the firing frequencies, together with the box plots of the range every 50,000 steps.}
    \label{fig:error:new}
\end{figure}
\subsection{Theoretical benefits}
\label{ssec:theoreical}
To discuss theoretical benefits, we will establish a fixed scenario for consistent analysis.
First, the root node of the spatial octree is located at level $1$ of the tree.
The branching level is $b$ as before.
We fix two neurons, $N$ and $N'$, at level $l \in O(\log{n})$ and assume that the tree is balanced, although our calculations work with inbalanced trees as well.
With each neuron, we associate a path ($P = (\text{root}, ..., N), P' = (\text{root}, ..., N')$) from the root node down the tree.
Neuron $N$ belongs to MPI rank $R$.
Lastly, neuron $N$ fires every $K$ simulation steps, while neuron $N'$ fires every $K'$ simulation steps.
\paragraph{Update of connectivity}
\label{par:connectivity}
We begin with the differences in connectivity updates, namely the Barnes--Hut algorithm and our location-aware Barnes--Hut algorithm.
For this, we must distinguish two scenarios:
First, assume $N$ proposes to a local neuron, i.e., $N'$ also belongs to MPI rank $R$ as well.
For this to occur, all nodes on $P'$ belong to $R$ as well.
Regardless of whether $N$ selected $N'$ in the first Barnes--Hut application or first chose inner nodes from $P'$ and then selected $N'$ later in a repeated application, neither version initiated a search on a node from a different MPI rank. 
Therefore, both versions perform identically.

Secondly, assume $N$ proposes to a distant neuron, i.e., $N'$ belongs to MPI rank $R' \not = R$.
For this to occur, on path $P'$, the first nodes up to the branching level $b$ belong to $R$, while all below belong to $R'$ (as indicated in Fig.~\ref{fig:tree:download}).
While the node at the branching level also belongs to $R'$, $R$ maintains a local copy.
If $N$ had not selected any inner nodes of $P'$ (again, because the acceptance criterion never permitted approximation at any level), both versions perform identically.
Any inner node selected above the branching level is irrelevant, as every MPI rank has a copy of the upper portion---no communication required.
In the remaining case, where $N$ selected inner nodes on the path from root to $N'$ at or below the branching level, our algorithms differ:
The original version would download all necessary information via RMA.
In contrast, our version transfers $N$'s information to $R'$ and receives a response.

What does this mean?
The length of $P'$ is $\text{level}(N')$, with $b$ nodes stored locally on $R$ and $\text{level}(N') - b$ nodes stored on $R'$.
In the worst case, this means $\text{level}(N') - b - 1 \in O(\log{n})$ searches starting at a foreign node---with required RMA communication---for the original version.
Our version sends $N$'s information to $R'$ immediately when an inner node from $P'$ belonging to $R'$ is selected.
The computation on $R'$ with $N$ is similar to what $R$ would compute, albeit with a different state in the pseudo-random number generator.
\edited{This, however, can be neglected as the PRNG state is inherently unknown, and thus, our algorithm produces the same qualitative results.}

Overall, our new algorithm reduces per-neuron communication from $O(log n)$ to $O(1)$ whenever the underlying assumptions hold. 
In pathological configurations, the complexity reverts to the old $O(log n)$ complexity.
Although we do not furnish a formal proof excluding such pathological cases, our experimental study demonstrates that they do not arise in practice. 
Hence, across all relevant scenarios, the effective communication cost is constant.
\paragraph{Update of electrical activity}
We compare direct spike communication in the original version with approximated firing frequencies.
We assume frequencies are updated every $\Delta$ simulation steps.
The analysis is straightforward:
First, we reduce the number of communications by a factor of $\Delta$.
Note that in the original version, each rank---even if it receives no incoming spikes---must obtain the number of incoming spikes (i.e., $0$ in this case).
Second, considering a period of $\Delta$ updates and assuming the synapse from $N$ to $N'$, the total communication in the original version is $O(K * 64\,\text{Byte})$ (it sends neuron IDs when they fire).
Furthermore, when determining if neuron $N$ spiked, $R'$ searches for $N$'s ID among all received IDs from $R$.
These are sorted, so this uses binary search.
In our proposed version, total communication is $O(64\,\text{Bytes})$ and instead of lookup, the rank generates a pseudo-random number.

Thus, our version can theoretically benefit from larger $\Delta$~values (i.e., less frequent updates as simulation runs longer and stabilizes).
Typical neurons fire at frequencies between 10~Hz and 100~Hz~\cite{fries2015rhythms}, corresponding to $K, K' \in [10, 100]$, since one simulation step equals 1~ms of biological time.
We choose $\Delta = 100$ (updating frequencies every time connectivity changes), and timings in Section~\ref{ssec:timings} show that---despite this being quite frequent---we still gain a significant advantage; even without theoretical benefits.
\subsection{Timings}
\label{ssec:timings}
We have performed our experiments on the Lichtenberg~2 Stage~2 High-Performance Cluster at Technical University of Darmstadt. 
Each of the relevant compute nodes has 512~GB DDR5-4800 main memory, two Intel Xeon Platinum 8470Q processors (52 cores, no hyperthreading, frequency of 2.1~GHz, 3.8~GHz with boost; AVX-512, AMX, VNNI, TSX-Ni).
The compute nodes are connected with InfiniBand HDR100 in a 1:1 non-blocking topology.
We used gcc-11.2.0, mpich-3.4.2, and openucx-1.12.0 on RedHat~8.10.
\edited{Note, however, that there are no other simulators capable of simulating structural plasticity at the size of our experiments besides the ones used, which hinders comparison across codes with, e.g., NEST.}

Our experimental setup remained consistent across all experiments:
We simulated 1000 update steps of the network (i.e., 10 plasticity updates), starting with no initial connectivity, and each neuron having between 1.1 and 1.5 vacant synaptic elements.
This ensures each neuron can connect to one other neuron and be connected to by one other neuron.
We used 10 plasticity updates because multiple neurons may select the same target (which accepts only one), requiring retries in subsequent updates.

\begin{figure*}
    \centering
    \includegraphics[width=1\linewidth]{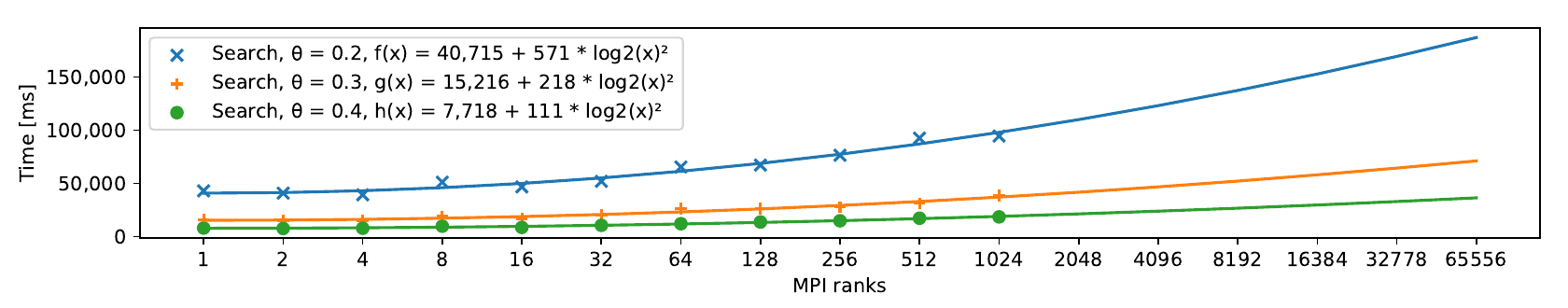}
    \caption{Timings for our new algorithm for 65,536 neurons per MPI rank, including performance models. \fabian{Note that we fitted the trend line and extrapolated it beyond our tests.}}
    \label{fig:extrapolation}
\end{figure*}

We ran our tests with \{1, 2, 4, 8, 16, 32, 64, 128, 256, 512, 1024\} MPI ranks, using \{1024, 4096, 16 384, 65 536\} neurons each, with Barnes--Hut acceptance criteria ($\theta$) of \{0.2, 0.3, 0.4\} (resulting in $11 \cdot 4 \cdot 3 = 132$ configurations); each configuration tested once with old algorithms and once with our improved algorithms.
We report average times for exchanging fired status and forming new synapses (including target finding, request exchange, request handling, and answer exchange).
We also report approximate byte counts sent, received, and remotely accessed by MPI ranks.
Here, ``approximate'' means we only count bytes we directly handle, not what the library communicates additionally.

\paragraph{Update of connectivity}
\label{par:timings:bh}
We present timings for both Barnes--Hut algorithms---the default one and our proposed one---in Figure~\ref{fig:timings:bh}.
It's clear that larger $\theta$ values (allowing more approximations) reduce connectivity update time.
This is expected.
Note that small MPI rank counts or neurons per MPI rank numbers mask noise-related issues we identify.
Timings for one MPI rank are essentially identical---as expected; our algorithm improves communication between MPI ranks.
Our algorithm's timings for ``small'' simulations (roughly up to 64 MPI ranks with 4096 neurons each) show no scaling because general synchronization for all-to-all communication and actual computation overwhelm the required communication.
Apart from this artifact, larger MPI rank counts yield greater benefits from our new algorithm, up to a 6x improvement for the largest simulation (or 10x improvement for 512 MPI ranks with 16,384 neurons each).
Our new algorithm also scales similarly to the old one with respect to $\theta$, with overall scaling behavior similar to the original algorithm---just significantly faster.
\edited{The same data, but shown for strong scaling, is visible in Fig.~\ref{fig:strong:bh}. This, however, must be taken with a grain of salt: When performing strong scaling in neuronal simulations, one usually uses hybrid parallelization (for example, using OpenMP threads), which does not show in the MPI communication.
In Fig.~\ref{fig:extrapolation}, we have used Extra-P\footnote{\url{https://github.com/extra-p/extrap/}} to extrapolate the timings of the new location-aware Barnes--Hut algorithm for 65,536 neurons per MPI rank. For the different approximation parameters, it gives a scaling behavior of $O(\log^2(n))$ with different coefficients.}
\paragraph{Update of electrical activity}
\label{par:timings:ea}
Figure~\ref{fig:timings:firing} displays times for both the default spike transfer method and our proposed frequency transfer approach.
With the old algorithm, we see slightly more than double the time to exchange firing IDs when doubling MPI ranks, and scaling that begins higher than quadrupling for a few MPI ranks but decreases to less than doubling for many MPI ranks when quadrupling neurons per rank.
This confirms we're on the right path---the synchronization and communication channel setup are the primary bottlenecks.
Spike lookup times after receipt remain essentially constant with MPI rank count---though slightly rising from 1 to 2 to 4 ranks (see Figure~\ref{fig:timings:distant}).
For our proposed algorithm, we observe a clear pattern:
Communication effort stays virtually constant with MPI rank count, scaling only roughly with neurons per rank.
However, times are many orders of magnitude lower than the old version (largest simulation: 23 s vs. 169 ms).
\edited{Figure~\ref{fig:strong:freq} shows the strong-scaling experiments.}

Looking up neuron frequencies and generating pseudo-random numbers takes more time than searching IDs, but not significantly more (9467 ms vs. 13 s), see Fig.~\ref{fig:timings:distant}.
From 64 MPI ranks upward, lookup times remain essentially constant.
Overall, this slowdown is a worthwhile trade-off for the speedup shown in Fig.~\ref{fig:timings:firing}.
\begin{figure*}
\centering
\begin{minipage}{.5\textwidth}
\captionsetup{width=.8\linewidth}
\centering
\includegraphics[width=\linewidth]{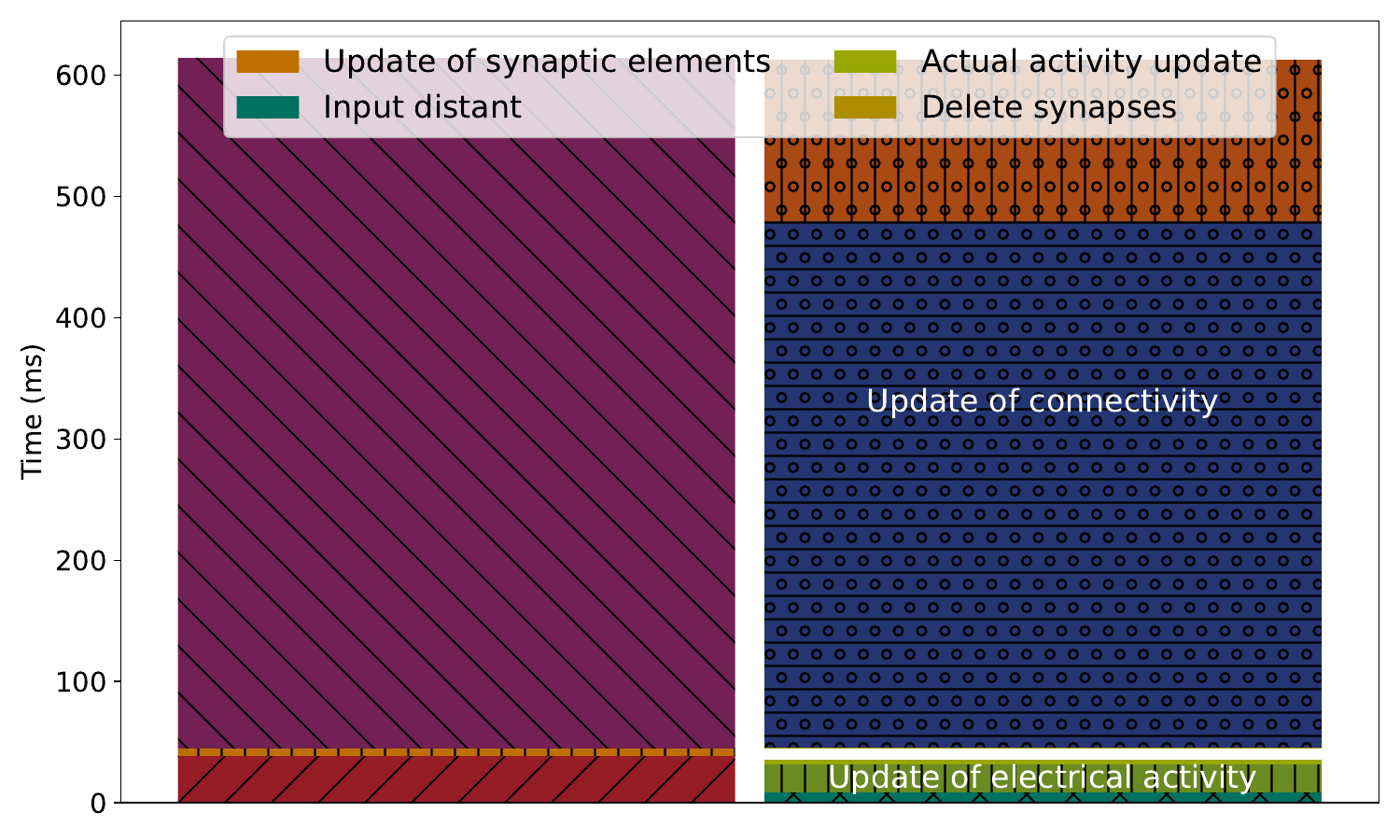}
\end{minipage}%
\begin{minipage}{.5\textwidth}
\captionsetup{width=.8\linewidth}
\centering
\includegraphics[width=\linewidth]{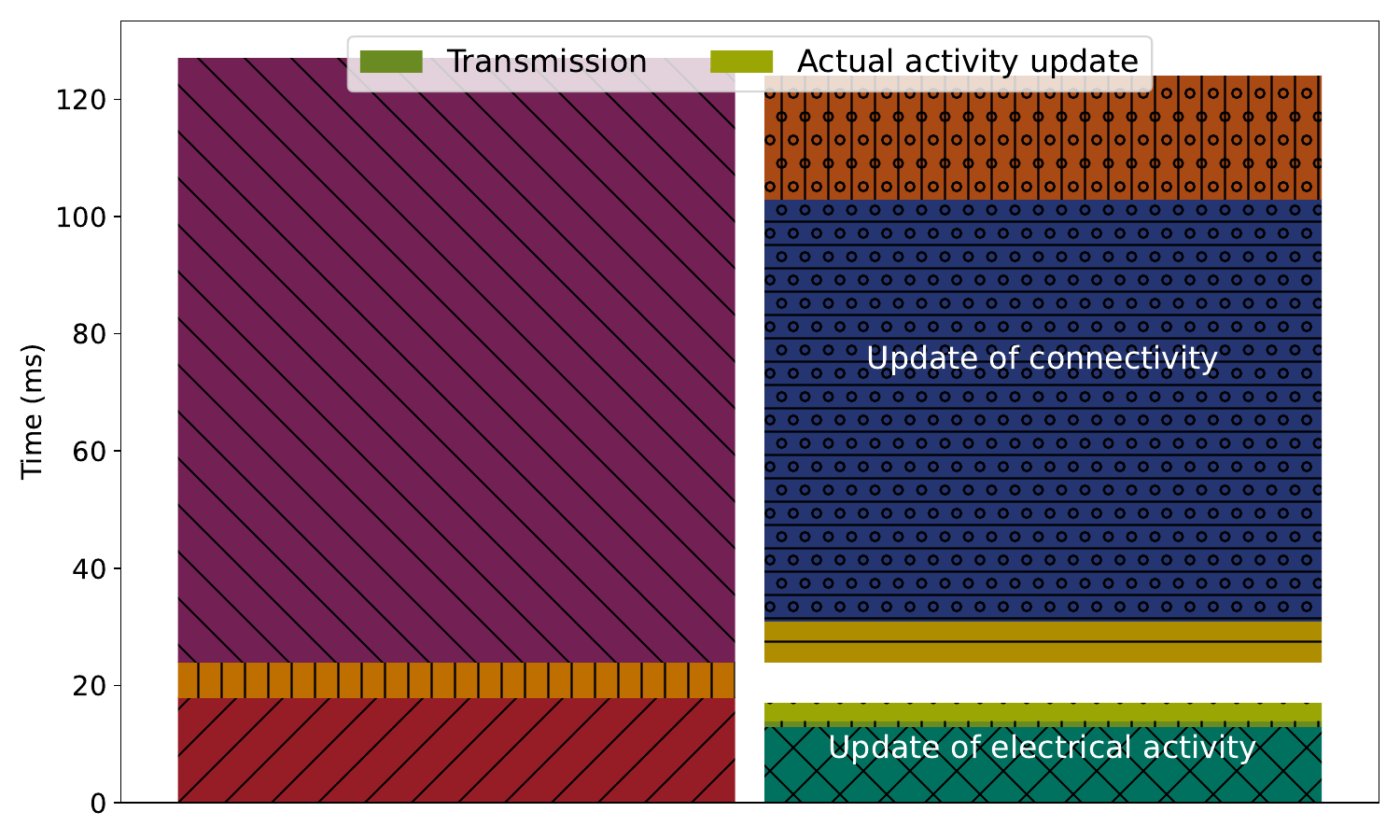}
\end{minipage}
\caption{Timings of the simulation with 1024 MPI ranks, 65,536 neurons per rank, and the approximation threshold $\theta = 0.2$. Note that for the timings for the old algorithm, ``Barnes--Hut'' includes remote memory accesses. Left: old algorithm. Right: proposed location-aware algorithm.}
\label{fig:runtime}
\end{figure*}
\subsection{Amount of communication}
\label{ssec:bytes}
Tables~\ref{tab:bytes:default} and~\ref{tab:bytes:new} show byte transfer amounts during simulations.
For the old Barnes--Hut algorithm and standard spike transfer, we list bytes sent and remotely accessed.
For our proposed location-aware Barnes--Hut algorithm and neuron spike approximation, we only list bytes sent, as no remotely-accessed bytes occur.
Technically, remotely-accessed bytes could happen, but this does not occur in practice since no MPI rank needs to fetch data.
In small simulations, our new algorithms transfer more bytes initially but quickly become superior.
As a direct result, we also reduce the memory footprint ``en passant'':
MPI ranks in the old algorithm had to download octree nodes from other ranks and cache them locally if different neurons needed the data during the same connectivity update---this is now nearly obsolete.
\begin{table}
\begin{tabular}{r|r|r|r|r}
& \multicolumn{4}{|c}{Neurons per rank} \\ 
MPI ranks & 1024 & 4096 & 16 384 & 65 536 \\ \hline\hline

\multirow{2}{3em}{1 r.} 
 & 86 KB & 324 KB & 1273 KB & 5075 KB \\ \cline{2-5}
 & 0 B & 0 B & 0 B & 0 B \\\hline\hline

\multirow{2}{3em}{2 r.} 
 & 845 KB & 3180 KB & 12 MB & 48 MB \\ \cline{2-5}
 & 3111 KB & 11 MB & 21 MB & 152 MB \\\hline\hline

\multirow{2}{3em}{4 r.} 
 & 2442 KB & 8921 KB & 34 MB & 136 MB \\ \cline{2-5}
 & 8878 KB & 53 MB & 189 MB & 392 MB \\\hline\hline

\multirow{2}{3em}{8 r.}  
 & 5956 KB & 20 MB & 78 MB & 312 MB \\ \cline{2-5}
 & 55 MB & 112 MB & 575 MB & 2087 MB \\\hline\hline

\multirow{2}{3em}{16 r.}  
 & 15 MB & 46 MB & 170 MB & 666 MB \\ \cline{2-5}
 & 166 MB & 510 MB & 1107 MB & 5576 MB \\\hline\hline

\multirow{2}{3em}{32 r.}  
 & 39 MB & 102 MB & 358 MB & 1379 MB \\ \cline{2-5}
 & 331 MB & 1311 MB & 4141 MB & 9 GB \\\hline\hline

\multirow{2}{3em}{64 r.}  
 & 108 MB & 237 MB & 755 MB & 2826 MB \\ \cline{2-5}
 & 1143 MB & 2362 MB & 9548 MB & 31 GB \\\hline\hline

\multirow{2}{3em}{128 r.}  
 & 437 MB & 698 MB & 1741 MB & 5912 MB \\ \cline{2-5}
 & 2739 MB & 7495 MB & 16 GB & 72 GB \\\hline\hline

\multirow{2}{3em}{256 r.}  
 & 1315 MB & 1862 MB & 3955 MB & 12 GB \\ \cline{2-5}
 & 5177 MB & 16 GB & 49 GB & 128 GB \\\hline\hline

\multirow{2}{3em}{512 r.}  
 & 3966 MB & 5577 MB & 9775 MB & 25 GB \\ \cline{2-5}
 & 14 GB & 29 GB & 106 GB & 357 GB \\\hline\hline

\multirow{2}{3em}{1024 r.} 
 & 17 GB & 24 GB & 32 GB & 65 GB \\ \cline{2-5}
 & 29 GB & 77 GB & 188 GB & 772 GB \\
 
\end{tabular}
\caption{Bytes sent and remotely-accessed for the old Barnes--Hut algorithm, the standard way to exchange the neuron spikes, and a small amount of bookkeeping. The values do not significantly differ when varying the approximation threshold $\theta$ (in the one-digit percent range), so we provide the values for $\theta = 0.2$. The upper values represent the total number of bytes sent/received, the lower values represent the total number of bytes remotely accessed. Digits after the decimal point are cut. 1~KB = 1024~B, etc.}
\label{tab:bytes:default}
\end{table}

\begin{table}
\begin{tabular}{r|r|r|r|r}
& \multicolumn{4}{|c}{Neurons per rank} \\ 
MPI ranks & 1024 & 4096 & 16 384 & 65 536 \\ \hline\hline

{1 r.} 
 & 211 KB & 814 KB & 3230 KB & 12 MB \\\hline
 
{2 r.} 
 & 757 KB & 2863 KB & 11 MB & 44 MB \\\hline

{4 r.} 
 & 1794 KB & 6914 KB & 26 MB & 107 MB \\\hline

{8 r.}  
 & 3925 KB & 14 MB & 58 MB & 233 MB \\\hline

{16 r.}  
 & 9908 KB & 32 MB & 123 MB & 486 MB \\\hline

{32 r.}  
 & 19 MB & 66 MB & 251 MB & 993 MB \\\hline

{64 r.}  
 & 41 MB & 134 MB & 509 MB & 2006 MB \\\hline

{128 r.}  
 & 192 MB & 380 MB & 1132 MB & 4140 MB \\\hline

{256 r.}  
 & 401 MB & 778 MB & 2286 MB & 8318 MB \\\hline

{512 r.}  
 & 867 MB & 1628 MB & 4648 MB & 16 GB \\\hline

{1024 r.} 
 & 8685 MB & 10 GB & 15 GB & 39 GB \\
 
\end{tabular}
\caption{Bytes sent for the proposed location-aware Barnes--Hut algorithm, the proposed way to exchange the neuron frequencies, and a small amount of bookkeeping. The values do not significantly differ when varying the approximation threshold $\theta$ (in the one-digit percent range), so we provide the values for $\theta = 0.2$. Digits after the decimal point are cut. 1~KB = 1024~B, etc.}
\label{tab:bytes:new}
\end{table}
\subsection{Quality of the approximation}
\label{ssec:quality}
For this test, we significantly modify the simulation setup.
First, we simulate only 32 neurons distributed across 32 MPI ranks (one neuron per rank)---this forces neurons to connect only to neurons on other ranks, fully utilizing our spike approximation.
Second, we simulate 200,000 steps with 2000 connectivity updates.
We set neurons' target activity to 0.7 and synaptic element growth rates to 0.001.
This causes neurons to seek 22--23 synapses.
Finally, we provide neurons with background activity from $\mathcal{N}(5,1)$ to sufficiently stimulate them for synaptic element growth (see~\cite{Butz2013} for a deeper discussion of this choice; same background activity in both simulations).

Figures~\ref{fig:error:old} and~\ref{fig:error:new} show calcium concentrations for both the standard spike transmission method and our proposed frequency-based spike approximation.
The initial phases are similar for both simulations:
After background activity raises neurons to approximately 0.4 calcium, they begin growing synapses.
They overshoot the target activity, prune some synapses, and then attempt to reach the target activity.
We observe the same fluctuations in both simulations.
\edited{Every 50,000 steps, we also plot the variation of calcium values to across the neurons. For both the old and the new algorithm, the median lies around 0.70--the target value. For the new algorithm, the algorithm has a larger inter-quartile range in steps 100,000 and 150,000---a lower one at step 200,000, but with an outlier. Overall, the results of our proposed algorithm are comparable to those of the old algorithm in terms of statistical variation.}
\subsection{Total time}
\label{ssec:total}
Figure~\ref{fig:runtime} shows timings for our longest simulation---1024 MPI ranks with 65,536 neurons per rank and $\theta = 0.2$.
The old algorithms for synapse formation and neuron spike transmission, plus the models for electrical activity, synaptic elements, and synapse deletion, took 617 s for 1000 steps.
Our new algorithms for synapse formation and neuron spike transmission, plus the unchanged models, took 131 s for 1000 steps.
This represents an absolute reduction of 486 s in wall-clock time, or approximately 78.8~\% relative reduction.
Examining the time distribution with our new algorithms reveals potential issues for even larger simulations:
With our proposed algorithm, 22 s of the 131 s (about 16~\%) involve straightforward per-neuron calculations (``Input distant'', ``Actual activity update'', ``Update of synaptic elements''), i.e., numerically solving differential equations or generating pseudo-random numbers.
28 s (about 21~\%) are spent during communication periods (``Synapse exchange'', ``Spike exchange'', ``Delete synapses'') (Note: no synapses were actually deleted; timing reflects just synchronization).
Meanwhile, we spent 72 s (about 55~\%) on Barnes--Hut computation, excluding remote memory accesses.
This part also occurs per-neuron basis, requiring no interactions with other neurons.
However, different neurons undergo vastly different Barnes--Hut computation paths (as two nearby neurons might choose targets in opposite brain regions).
This contrasts with the earlier-mentioned per-neuron work, which is identical for every neuron.

\section{Conclusion}
\label{sec:conclusion}
We have presented two algorithms to improve MSP-based simulation runtimes, which can predict brain changes after learning, lesions, or normal development.
Building on the Barnes--Hut algorithm and an earlier 2018 algorithm~\cite{Rinke2018}, we improve connectivity update time by a factor of six, neuron spike transmission time by more than 100 times, and transferred information amount by a factor of 21.
We incur a 1.5 times increase in neuron spike lookups, an insignificant cost compared to the gains.
We also proved a theoretical communication complexity improvement from logarithmic to constant.
Overall, our algorithms reduce the wall-clock time of the largest simulation by approximately 78.8~\%.

\edited{
With our work, simulating the entire human brain becomes feasible.
For instance, Fugaku consists of 158,976 nodes with 48 cores each.
With 65,536 neurons per core, we require 32k compute nodes.
This enables neuroscientists to simulate structural plasticity on a whole-brain level.
As structural plasticity plays an important role in learning and development, as well as in degenerative diseases such as Alzheimer's and Parkinson's, our approach can help find therapies or cures.
For more realistic models, it could be possible to integrate structural connectomes, e.g., retrieved from polarized light images.
Currently, there is no model available representing the entire brain, but technical advances will hopefully enable higher resolutions and larger tissues to make a complex model possible in the future.
In the future, GPUs can lower the computational need even more, making it more feasible for everyday simulations, enabling prediction and individual treatment plans for individual patients.
}

\fabian{Our work readily generalizes to other problem domains with similar structure: Firstly, when the actual data can be approximated by its ``shape'' (as in: the spikes can be approximated by their frequency), one can substitute multiple rounds of communication with a single round of communication (with more data) and local computation, reducing the time it takes to synchronize between processes. Secondly, when an algorithm requires that processes use distributed data in their computation (as in: the Barnes--Hut algorithm needs nodes stored on other processes), it is often intuitive to use RMA. As we showed, however, two rounds of synchronous communication sandwiching local computation can be a worthwhile alternative, reducing both synchronization and waiting times.} 

\section*{Acknowledgements}
This work was funded by the Deutsche Forschungsgemeinschaft (DFG, German Research Foundation) – Project No. 449683531 (ExtraNoise).
The authors gratefully acknowledge the Federal Ministry of Research, Technology and Space (BMFTR) and the Hessian Ministry of Science and Research, Art and Culture (HMWK) for supporting this work as part of the NHR funding. 
The authors gratefully acknowledge the computing time provided to them on the high-performance computer Lichtenberg II at TU Darmstadt, funded by the German Federal Ministry of Research, Technology and Space (BMFTR) and the State of Hesse.
The authors would like to thank the NHR-Verein e.V. for supporting this work within the NHR Graduate School of National High Performance Computing (NHR).


\FloatBarrier

\bibliographystyle{ieeetr}
\bibliography{references.bib}

\end{document}